\documentstyle[emulateapj]{article}

\input psfig.sty
\submitted{To appear in ApJ 2001 April 10}
\begin{document}

\newcommand{\bl}[1]{\mbox{\boldmath$ #1 $}}
\newcommand{\cs}{c_{\rm s}}
\newcommand{\cmc}{{\rm cm}^{-3}}
\newcommand{\Alf}{Alfv\'{e}n\ }
\newcommand{\Gammat}{\Gamma_{\rm turb}}
\newcommand{\Gammacr}{\Gamma_{\rm cr}}
\newcommand{\td}{t_{\rm d}}
\newcommand{\tc}{t_{\rm c}}
\newcommand{\tff}{t_{\rm ff}}
\newcommand{\Mdotw}{\dot{M}_{\rm w}}
\newcommand{\Mdotsf}{\dot{M}_{\rm SF}}
\newcommand{\vw}{v_{\rm w}}
\newcommand{\vrms}{v_{\rm rms}}
\newcommand{\cm}{\rm cm}
\newcommand{\kms}{\rm km~s^{-1}}
\newcommand{\s}{\rm s}
\newcommand{\pc}{\rm pc}
\newcommand{\erg}{\rm erg}
\newcommand{\K}{\rm K}
\newcommand{\bfr}{\bf r}
\newcommand{\bfS}{\bf S}
\newcommand{\bfOmega}{\bf\Omega}
\newcommand{\LCO}{L_{\rm CO}}
\newcommand{\Lbol}{L_{\rm bol}}

\title{Constraints on Stirring and Dissipation of MHD Turbulence \\
        in Molecular Clouds}
\author{Shantanu Basu}
\affil{Department of Physics and Astronomy, University of Western Ontario,
London, Ontario N6A 3K7, Canada; basu@astro.uwo.ca.}
\centerline{\sc and}
\author{Chigurupati Murali}
\affil{Department of Astronomy, University of Massachusetts, Amherst,
MA 01003-4525, USA; murali@kea.astro.umass.edu.}

\begin{abstract}
We discuss constraints on the rates of stirring and dissipation of MHD
turbulence in molecular clouds.  Recent MHD simulations suggest that
turbulence in clouds decays rapidly, thus providing a significant
source of energy input, particularly if driven at small scales by,
for example, bipolar outflows. We quantify the heating rates by combining 
the linewidth-size relations, which describe global cloud properties, 
with numerically determined dissipation rates.  We argue
that, if cloud turbulence is driven on small internal scales, the $^{12}$CO 
flux (enhanced by emission from weakly supersonic shocks) will be much larger 
than observed; this, in turn, would imply excitation temperatures 
significantly above observed values. We reach two conclusions: 
(1) small-scale driving by bipolar outflows cannot possibly account
for cloud support and yield long-lived clouds, unless the published MHD
dissipation rates are seriously overestimated; (2) driving on large 
scales (comparable to the cloud size) is much more viable from an energetic
standpoint, and if the actual net dissipation rate is only slightly lower than 
what current MHD simulations estimate, then the observationally inferred 
lifetimes and apparent virial equilibrium of molecular clouds can be 
explained.

\end{abstract}


\section{Introduction}
Millimeter-wave surveys of CO line emission have established the
large-scale distribution and properties of molecular clouds in the
Galaxy (Solomon, Sanders, \& Scoville 1979; Leung, Kutner, \& Mead
1982; Dame et al. 1986; Solomon et al. 1987). These surveys reveal two
essential features of clouds: (1) supersonic linewidths, with observed
velocity dispersion $\sigma$ exceeding the isothermal sound speed
($\cs \approx 0.2$ km s$^{-1}$, corresponding to the measured
temperature $T \approx 10\,\K$) by factors $\sim 10$ or more; and (2)
power-law relations between $\sigma$ and the cloud size $L$,
of the approximate form $\sigma \propto L^{0.5}$, and between the mean
density $n$ and $L$, of the approximate form $n \propto L^{-1}$
(hereafter collectively labeled the {\it linewidth-size-density} relations).  
The magnitude of the observed linewidths is essentially that required to
put individual clouds in virial equilibrium, given the estimated sizes
and masses (e.g., Larson 1981; Myers 1983; Solomon et al. 1987).

Turbulent motions are often invoked as an origin for the supersonic
linewidths since they can provide an effective pressure (Chandrasekhar
1951) which, unlike the much lower thermal pressure, is adequate to
provide cloud support. However, supersonic hydrodynamic turbulence
should decay rapidly through shocks (e.g., Mestel 1965; Goldreich \&
Kwan 1974). Since molecular clouds tend to have large scale magnetic
fields (see Crutcher 1999), an attractive possibility has been that
the turbulence represents supersonic but sub-Alfv\'{e}nic
magnetohydrodynamic (MHD) waves, of which the non-compressive shear
\Alf mode may be a long-lived component (Arons \& Max 1975;
Mouschovias 1975).

However, recent numerical MHD simulations suggest that a spectrum of
MHD waves, or MHD turbulence, cannot persist in a cloud for many
crossing times (e.g., Stone, Gammie, \& Ostriker 1998; Mac Low et
al. 1998).  These simulations tend to show that compressive motions
and shocks arise so readily that the turbulence decays in one (or even
much less than one) turbulent crossing time of the cloud
($\tc=L/\sigma$).  This result is particularly surprising in the
simulations of Stone et al. (1998), who stir the system only through
incompressible modes ($\nabla \cdot\delta{\bl v}=0$).  If
confirmed, these results imply that molecular cloud turbulence must be
constantly and strongly driven if clouds survive for up to several crossing
times, as implied by the inferred virial equilibrium of the clouds
and their estimated lifetimes (e.g., Blitz \& Shu 1980; Williams \& McKee 1997).

But are such high rates of dissipation really allowed, even if a
steady-state is maintained by equally strong stirring?  In this 
paper, we consider this question by examining the energetics of
clouds with turbulent dissipation.  
In \S\ 2, we determine heating
rates in clouds obeying the linewidth-size-density relations based on
dissipation rates determined from numerical simulations.  In \S\ 3,
the corresponding luminosities are determined and compared with
observed luminosities. Further constraints on stirring are given in \S\ 4,
and our conclusions are presented in \S\ 5.

\section{Heating Rates}
Define $\Gamma({\bl r})$ as the heating rate per unit volume as a
function of position ${\bl r}$ in a molecular cloud. Various processes can
contribute to $\Gamma({\bl r})$.  Thermal energy input from cosmic rays
at the mean rate
\begin{equation}
\Gammacr=6.4\times 10^{-25} \, (n/10^3 \, \cm^{-3}) \; \erg \, \cm^{-3}\,
\s^{-1}
\label{eq:cr}
\end{equation}
has been thought to maintain clouds of mean number density $n$ at
temperatures $T\sim 10\,\K$ (e.g., Goldsmith \& Langer 1978).

A second contribution could also come from the decay of the MHD turbulence
that is thought to support molecular clouds globally.  The turbulent
heating rate $\Gammat$ can be written as the ratio of the energy
density $U$ of the cloud to an arbitrary dissipation timescale $\td$:
$\Gammat=U/\td$.  The results of recent numerical simulations (Stone
et al. 1998; Mac Low 1999) suggest that the ratio of dissipation time
to turbulent crossing time $\kappa\equiv \td/\tc \leq 1$ under typical
conditions in molecular clouds.

For a cloud with mean density $\rho$ and one-dimensional non-thermal
velocity dispersion $\sigma$, the turbulent energy density is $U=3/2
\rho \sigma^2$, so that the turbulent heating rate
\begin{equation}
\Gammat = \frac{3}{2} \frac{\rho \sigma^2}{\td} 
= \frac{3}{2} \, \kappa^{-1} \frac{\rho \sigma^3}{L}.
\label{eq:dissip1}
\end{equation}

There are a variety of other processes which can contribute to the
energy balance in molecular clouds.  They include the Galactic ultraviolet
background which heats the edges of clouds and the internal radiation
field generated by the stars which form in localized regions of clouds.
Here we neglect processes which might contribute only to localized
regions of star formation and cloud boundaries.

\subsection{Stirring Scale}
An important property of the turbulent dissipation rate is its
dependence on stirring scale.  Dimensionally, the dissipation rate
\begin{equation}
\Gammat=\eta \, {\rho\sigma^3\over \lambda},
\label{eq:dissip4}
\end{equation}
where $\lambda$ is the driving scale and $\eta$ is, in general, a
dimensionless function of $\rho$, $\sigma$ and $\lambda$. Kolmogorov
and Obukhov (e.g., Landau \& Lifshitz 1987) originally derived the
value of $\eta$ in a self-similar cascade which is appropriate to the
decay of incompressible hydrodynamic turbulence.

By analogy, Stone et al. (1998) and Mac Low (1999) derive values of
$\eta$ for their simulations of compressible MHD turbulence.  In this
case, $\rho$ is the mean density of the computational region. Their
results indicate that $\eta\approx 1$ over a range of driving
scales\footnote{Mac Low (1999) defines a constant $\eta_v$ using the
three-dimensional velocity dispersion and other numerical factors, so that our
$\eta=32.6\, \eta_v$.}, implying that $\td$ is of order the crossing time
$\tc$ or less.  In particular, comparing equations (\ref{eq:dissip1})
and (\ref{eq:dissip4}) shows that
\begin{equation}
\kappa = \frac{3}{2}\,  \eta^{-1} \, \frac{\lambda}{L}.
\end{equation}
Since driving may occur on scales $\lambda$ much smaller than typical
cloud sizes $L \sim {\rm few} - 100$ pc (e.g., through outflows), the
results of numerical simulations suggest that $\kappa$ may be much
less than unity.  This can obviously result in very large cloud
luminosities; below we determine the expected fluxes for these heating
rates.

\subsection{Dissipation in Clouds}
The linewidth-size relation is an empirically derived scaling between
the velocity width of $^{12}$CO ($J=1-0$) lines observed in a particular
molecular cloud and the size of that cloud.  Clouds approximately obey
the scaling determined by Solomon et al. (1987):
\begin{equation}
\sigma=0.72 \, (R/\pc)^{0.5} \; \kms,
\label{eq:lwsize}
\end{equation}
where $R\approx{1\over 2} L$ is an effective radius of the cloud.  The
same study also established an approximate density-size relation
\begin{equation}
\rho=134 \, (R/\pc)^{-1} \; M_{\odot} \, \pc^{-3}.
\label{eq:rhosize}
\end{equation}
These relations imply that the most massive molecular clouds are
approximately gravitationally bound and in virial equilibrium.
Although clouds have very complicated density structures, these
relations give a reasonable estimate of the mean density, mean
turbulent energy, and size of a cloud.  Thus, they allow an
estimate of the bolometric luminosity of a cloud, given a
mean dissipation rate as determined, say, from numerical simulations.

By combining the linewidth-size and density-size relations, we obtain
the mean heating rate due to dissipation of energy in turbulence:
\begin{eqnarray}
\label{eq:dissip2}
\Gammat &=& {7.1\times 10^{-11} \: \td^{-1} \; \erg \, \cm^{-3} \,\s^{-1}} \\
	&=& 5.4 \times 10^{-25} \, \kappa^{-1} \, (n/10^3 \,\cm^{-3})^{1/2} \; 
	\erg \, \cm^{-3} \, \s^{-1}. \nonumber
\end{eqnarray}
This relation should hold roughly for clouds with mean density in the
range $10^2\, \cmc \leq n \leq 10^4\, \cmc$.  Also note that the
energy density (or pressure) is independent of cloud size.

\section{Cloud Luminosities}
Assume only that the cloud exists in some definite volume and that it
is in a steady-state equilibrium.  Under these circumstances,
the total cooling rate within the volume must balance the total
heating rate.  Therefore, although the cloud may have a rather complex
density structure, and a variety of atomic and molecular species
(each having a different optical depth) may contribute to the cooling, the
bolometric luminosity $\Lbol$ must balance the volume-integrated heating
rate:
\begin{equation}
\label{eq:volint}
\dot U=\int d{\bl r} \Gamma({\bl r})=\Lbol.
\end{equation}
Note that we have neglected the possibility that mass loss can carry
away excess kinetic energy from the cloud in the form of a global wind
or outflow.

Of course, determining the bolometric luminosity of a cloud is no
trivial task.  Current observations usually only consider specific
atomic and molecular line transitions, with $^{12}$CO ($J=1-0$) by far the
most thoroughly investigated.  In their study, Solomon et al. (1987)
determine a $^{12}$CO ($J=1-0$) luminosity-linewidth relation for the clouds
in their sample (their eq. [9]). We use the linewidth-size-density relations
and convert units in their formula to express the empirical $^{12}$CO 
($J=1-0$) luminosity as
\begin{equation}
\LCO'=3.1\times 10^{30}\, (n/10^3\,\cmc)^{-5/2}\, \erg \, \s^{-1}.
\end{equation}
This luminosity represents only the flux through the portion of the
cloud's surface which is visible to the observer. Here we adopt a total
CO luminosity $\LCO=4\LCO'$, appropriate for a spherical cloud.
Of course, real molecular clouds have irregular shapes and surfaces, so that
the multiplicative constant may be somewhat greater than 4, but it should  
be no more than an order unity correction.

For clouds whose average properties obey the linewidth-size-density
relations, the bolometric luminosity determined from equation 
(\ref{eq:volint}) is 
\begin{equation}
\Lbol = 8.5\times 10^{32}\, \left[1.2\,(n/10^3\, )^{-2} + 
	\kappa^{-1}\, (n/10^3\, )^{-5/2}\,\right]\,  \erg \, \s^{-1},
\label{eq:engen}
\end{equation}
where the first term in brackets gives the cosmic ray heating from
equation (\ref{eq:cr}) and the second term gives the turbulent heating
from equation (\ref{eq:dissip2}).

\bigskip

\psfig{file=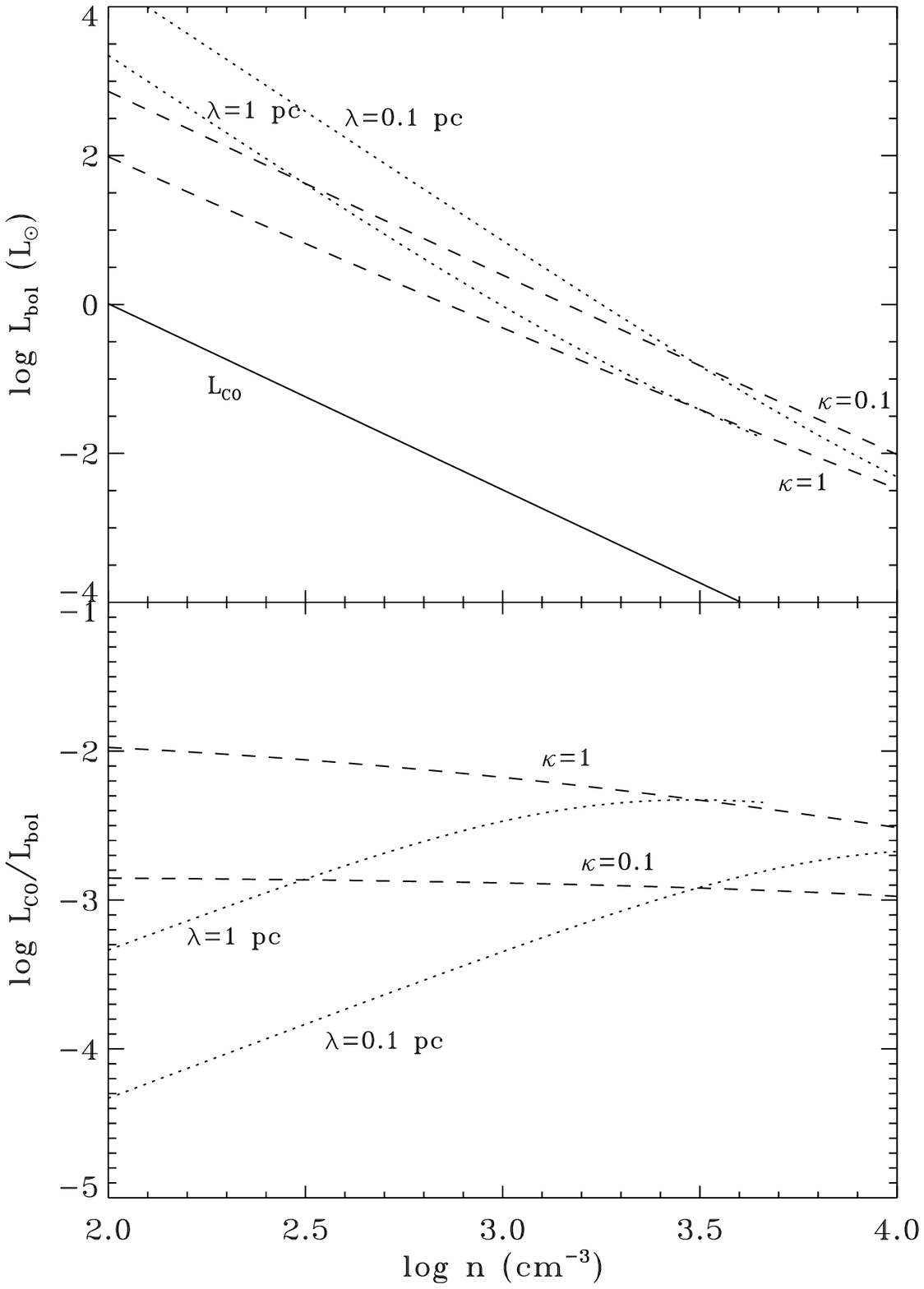,width=8.5cm}
\vspace{0.18in}
\smallskip\noindent {\sl Fig. ~1~}\ Top: Bolometric luminosity $\Lbol$
(equal to the sum of cosmic ray heating and turbulent dissipation)
as a function of mean density $n$ for clouds which obey the
linewidth-size-density relation and have fixed values of 
the dissipation parameter $\kappa$ (dashed lines) or the 
turbulent driving scale $\lambda$ (dotted lines).
The solid line shows the
$^{12}$CO luminosity $\LCO$ from the empirical relation determined
by Solomon et al. (1987).  Bottom: Ratio of observed $\LCO$ to bolometric
luminosity $\Lbol$ for the different values of $\kappa$ and $\lambda$.

\bigskip

The top panel of figure 1 shows the bolometric luminosity 
(equivalent to the heating rate)
compared with the expected $\LCO$
from Solomon et al. (1987). Dashed lines show $\Lbol$ for two values of 
$\kappa$,
and dotted lines show $\Lbol$ for two values of a fixed driving scale $\lambda$
that are compatible with driving by bipolar outflows. The bottom panel 
shows the flux ratio $\LCO/\Lbol$, thereby illustrating
how much emission is required beyond the $^{12}$CO
$J=1-0$ transition to balance the heating rate.  The excess can reach
(or even exceed) $3-4$ orders of magnitude, depending on the cloud mean 
density and driving scale. Detailed models of molecular line cooling
(Neufeld, Lepp, \& Melnick  1995) suggest that $\LCO/\Lbol$
typically falls in the range $1\%-2\%$. Therefore, it may be difficult to 
account for an additional $1-2$ orders of magnitude excess emission. 
Yet this is required if clouds are stirred from within, since bipolar outflows
are usually invoked as the only viable stirring mechanism (Norman \& Silk 1980; 
McKee 1989; Shu et al. 1999),
and these act on relatively small scales $\lambda\approx 0.1-1\, \pc$. 

Can the required excess emission be hidden in the high $J$ transitions of CO,
or in lines from other species?
In the recent simulations of Stone et al. (1998),
Smith, Mac Low, \& Zuev (2000; hereafter SMZ), and Smith, Mac Low, \& Heitsch 
(2000; hereafter SMH),
some $50\%-68\%$ of the turbulent energy dissipates
in shocks.  In simulations of freely decaying MHD turbulence, SMZ 
find that most shocks are slow, with typical Mach number in the range $1-3$,
i.e., velocities in the range $v_s = 0.2-0.6$ km s$^{-1}$ for a 
kinetic temperature $T=10\, \K$.
Applying the Rankine-Hugoniot jump conditions for a nonradiative shock, this
range of $v_s$ yields 
post-shock 
temperatures at best as high as $T=27\, \K$ (for
$v_s = 0.6$ km s$^{-1}$). SMH find only slightly higher Mach number velocity
jumps for driven MHD turbulence in systems with an energetically significant 
magnetic field.  Even if shock velocities are as high as
$v_s = 1$ km s$^{-1}$, the maximum possible (nonradiative)
post-shock temperature is $T=87\, \K$, which lies within the range in
which emission from CO dominates the molecular emission for $n \lesssim 10^4$
cm$^{-3}$ (see Fig. 4 of Neufeld et al. 1995). 
Furthermore, the low order transitions
of CO are easily excited under these conditions, with the lowest transition
of $^{12}$CO ($J=1-0$) accounting for $\LCO \sim 10^{-2}\Lbol$.
(Neufeld et al. 1995). Therefore,
the high bolometric luminosities $\Lbol$ shown in Figure 1 for small 
scale driving (implying an observed $\LCO/\Lbol \sim 10^{-4}-10^{-3}$) are 
incompatible with molecular cooling models. 
It is likely that a detailed calculation
based on the distribution of shock strengths in a typical simulation 
with small-scale driving will
show a significant excess of $\LCO$ above the empirical values presented 
in Figure 1. This will also imply much greater values of the 
excitation temperature than actually observed.

Finally, we note the scalings of the luminosity and energy generation rate.
Since cloud luminosities scale as $n^{-2.5}$ (alternatively $L^{2.5}$,
leading to the mass-luminosity scaling $M \propto L^{0.8}$ measured by
Solomon et al. 1987), it may be difficult to attribute their radiative
output to cosmic ray heating, which on average scales as $n^{-2}$ (or
$L^2$). Turbulent dissipation proportional to the crossing time
provides the correct scaling for energy input, i.e., when the driving
scale is proportional to the individual cloud size $L$ rather than a
fixed internal scale $\lambda < L$.
In fact, dissipation with $\kappa \gtrsim 1$, consistent with 
$\lambda \approx L$, 
yields $\LCO/\Lbol \sim 10^{-2}$ (see Fig. 1), roughly compatible 
with the observed $\LCO$ and the application of molecular cooling models
to post-shock conditions of weakly supersonic shocks.

\section{Other Constraints}

In addition to the luminosity
problem, another feature of rapid internal driving is that 
an amount of energy in excess of the
gravitational binding energy of a cloud needs to be input and
dissipated every crossing time (since $\lambda \ll L$). This is not likely
to be a very stable situation, as any fractional imbalance between dissipation
and cooling can lead to expansion and mass flow from the cloud.

Furthermore, the star formation rate
required to support this energy input also appears to be too large. An
estimate for the required star formation rate can be obtained in the
manner first done by McKee (1989) (see also Shu et al. 1999), but
incorporating the published dissipation rates of Stone et al. (1998)
and MacLow (1999), and using the linewidth-size-density relations.
The local dissipation rate given by equation (\ref{eq:dissip4}), with
$\eta = 1$, yields a Galactic dissipation rate
\begin{equation}
L_{\rm turb} = \frac{M\, \sigma^3}{\lambda},
\label{eq:dissip5}
\end{equation}
where $M$ is the mass of molecular gas in the Galaxy. We compare this
to the input rate from momentum-driven bipolar outflows with launching
speed $\vw$ and mass flux $\Mdotw$ which equals a fraction $f$ of the
star formation accretion rate $\Mdotsf$. If the swept-up shell
effectively dissipates when the speed drops to $\vrms = \sqrt{3}
\sigma$, the delivered energy is
\begin{equation}
L_{\rm out} = \frac{1}{2} f \Mdotsf \vw \vrms,
\label{eq:outflow}
\end{equation}
Equating equations (\ref{eq:dissip5})
and (\ref{eq:outflow}), the required star formation rate is
\begin{equation}
\Mdotsf = \frac{2}{3^{1/2} f} \; \frac{M \, \sigma^2}{\lambda \, \vw}.
\label{eq:sfrate}
\end{equation}
To obtain a numerical estimate, we use $f=1/2$ (e.g., Shu et
al. 1999), $\vw = 200$ km s$^{-1}$, and a relatively large driving
scale $\lambda = 1$ pc for the outflows. The total mass $M \simeq 10^9
M_{\odot}$ of molecular gas in the Galaxy is concentrated in the giant
molecular clouds (e.g., Solomon et al. 1987), for which we take 
$n=10^2 \, \cmc$ to be a
representative density. Equations (\ref{eq:lwsize}) and
(\ref{eq:rhosize}) then yield $\sigma = 3.48$ km s$^{-1}$. 
Finally, evaluating equation (\ref{eq:sfrate}) yields 
$\Mdotsf = 143 \, M_{\odot} {\rm yr}^{-1}$,
which is clearly too high compared to most estimates of the Galactic
star formation rate $\approx 3-5 \, M_{\odot} {\rm yr}^{-1}$ (see McKee 1989).
Altogether, recovering a low enough star formation rate given the fast
dissipation rates can be done only with extreme fine tuning, and 
therefore seems unlikely. Our conclusion is similar to that of McKee (1989)
that bipolar driving is adequate if the dissipation occurs over
many free-fall times $\tff$ of the cloud, but is inadequate if it occurs
over $\sim 1 \tff$.

\section{Conclusions}
In molecular clouds, energy generation by turbulent dissipation can
far exceed heating by cosmic rays (c.f. eq. [\ref{eq:engen}]), which have
traditionally been viewed as the energy source responsible for the
$10\, \K$ excitation temperatures measured in $^{12}$CO (Goldsmith \& Langer 
1978; Neufeld et al. 1995).  In this paper, we have demonstrated that 
rapid internal turbulent driving on small scales, in clouds which obey the 
linewidth-size-density relation, and in which MHD turbulence damps on 
essentially a crossing time over the driving scale, requires that the 
observed $^{12}$CO ($J=1-0$) luminosity $\LCO$ undersample the bolometric cloud 
luminosity $\Lbol$ by a far greater factor than predicted in standard molecular 
cooling models (Neufeld et al. 1995). 
We are left with two distinct possibilities: either current numerical
simulations are seriously overestimating the dissipation rate of MHD
turbulence, or the idea of
long-lived molecular clouds supported for several global crossing times by
small-scale internal driving, is incorrect.

If stirring occurs on approximately the scale $L$ of each individual
cloud, then the observed magnitude and scaling of $\LCO$ can be understood
with MHD dissipation rates comparable to those calculated in recent 
simulations. This may mean that clouds are not long-lived and need not
maintain a steady-state through stirring equal to the dissipation rate,
consistent with the conclusion (based on stellar ages) that star formation
occurs on a single cloud crossing time (Elmegreen 2000).
Alternatively, clouds need to survive for only $\approx 2-3$ 
crossing times to explain their estimated lifetimes and their apparent
state of virial equilibrium. We note that the crossing time for clouds which
obey the linewidth-size-density relation is 
\begin{equation}
\tc = L/\sigma = 1.3 \times 10^7 \,(n/10^2 \cm^{-3})^{-1/2}\; {\rm yr},
\end{equation}
which is $\approx 4$ times the free-fall time 
\begin{equation}
\tff = (3\pi/32G\rho)^{1/2} = 3.3 \times 10^6 \,(n/10^2 \cm^{-3})^{-1/2} \; {\rm yr}.
\end{equation}
The calculated $\tc$ is only a factor $\approx 2-3$ lower than the
estimated lifetime $\approx 3 \times 10^7$ yr of giant molecular clouds
(Williams \& McKee 1997) of mean density $n \approx 10^2\, \cmc$.
Thus, we believe that future numerical simulations which yield
slightly (a factor $\approx 2-3$) lower net dissipation rates can
reconcile turbulent dissipation (if driving occurs on the scale $L$ of 
each cloud) with the apparent evidence for long-lived 
(lifetime $\approx 2-3\, \tc
\sim 10 \, \tff$) clouds that are in near virial equilibrium.
These somewhat lower dissipation rates may be obtained by tapping
previously unmodeled global sources of turbulence such as gravitational
contraction and participation in the differential rotation of the Galaxy,
and/or dealing more carefully with heating and cooling of low density
molecular cloud envelopes. The latter requires dropping the isothermal
assumption, which can overestimate radiative losses and enhance the
development of shocks, particularly in low density regions.

\begin{acknowledgments}
We thank Ted Bergin for discussions and encouragement, and an anonymous 
referee for insightful comments. SB was supported by a grant from the 
Natural Sciences and Engineering Research Council of Canada.
\end{acknowledgments}



t
b
l

\end{document}